# Acoustic and Optical Phonon Frequencies and Acoustic Phonon Velocities in Silicon-Doped Aluminum Nitride Thin Films

**Dylan Wright[1,2], Dinusha Herath Mudiyanselage[3], Erick Guzman[1,2], Xuke Fu[1], Jordan Teeter[1,2], Bingcheng Da[3], Fariborz Kargar[4,*], Houqiang Fu[3], and Alexander A. Balandin[1,2,*]**

[1]Department of Materials Science and Engineering, University of California, Los Angeles, California 90095 USA

[2]California NanoSystems Institute, University of California, Los Angeles, California 90095 USA

[3] School of Electrical, Computer, and Energy Engineering, Arizona State University, Tempe, Arizona 85287, USA

[4]Materials Research and Education Center, Department of Mechanical Engineering, Auburn University, Auburn, Alabama 36849 USA

## Abstract

We report the results of the study of the acoustic and optical phonons in Si-doped AlN thin films grown by metalorganic chemical vapor deposition on sapphire substrates. The Brillouin – Mandelstam and Raman light scattering spectroscopies were used to measure the acoustic and optical phonon frequencies close to the Brillouin zone center. The optical phonon frequencies reveal *non-monotonic* changes, reflective of the variations in the thin-film strain and dislocation densities with the addition of Si dopant atoms. The acoustic phonon velocity decreases *monotonically* with increasing Si dopant concentration, reducing by ~300 m/s at the doping level of $3 \times 10^{19}$ cm$^{-3}$. Knowledge of the acoustic phonon velocities can be used to optimize ultra-wide bandgap semiconductor heterostructures and minimize the thermal boundary resistance of high-power devices.

---

[*] Corresponding authors: fkargar@auburn.edu (F.K.); balandin@seas.ucla.edu (A.A.B.)

Ultra-wide bandgap (UWBG) semiconductors constitute an increasingly important class of materials owing to their applications in power electronics and UV optoelectronics.[1–6] Conventionally, the UWBG group includes materials with band gaps above that of GaN, *i.e.* ≥3.4 eV. Among these materials, aluminum nitride (AlN) has attracted attention for its large electronic bandgap of ~6.2 eV, high thermal conductivity, with bulk values reaching above ~300 W/(mK),[7–10] and convenient *n*- and *p*-type doping.[11] For $Al_xGa_{1-x}N$ (0≤ x ≤ 1), silicon (Si) is a promising *n*-type dopant, which forms shallow donor states for x < 0.8.[12] For AlN-based devices, one needs to use controllable Si doping in a wide range from $10^{17}$ $cm^{-3}$ to $10^{20}$ $cm^{-3}$.[13,14] This range of silicon doping offers the required carrier concentrations and mobilities for practical device applications. On the other side, dopant atoms may act as the scattering centers for electrons and acoustic phonons, reducing the mobility and thermal conductivity. Dopant introduction can also change the distribution of stress and strain in AlN films and lead to the formation of dislocation line defects.[12]

We have recently found that boron (B) doping of diamond, another important UWBG material, results in substantial variation of the bulk and surface acoustic phonon frequency and velocity.[15] The change in the velocity of the shear-horizontal and the pseudo-longitudinal surface acoustic phonons in the degenerately doped diamond, as compared to that in the undoped diamond, was as large as ~15% and ~12%, respectively. The velocity of the bulk longitudinal acoustic (LA) and transverse acoustic (TA) phonons also decreased. The decrease in the acoustic phonon velocity has important implications for thermal transport.[16,17] Acoustic phonons are the main heat carriers in UWBG materials. The phonon scattering rate on dopant atoms, *via* the point-defect mechanism, is given by $\tau_P^{-1} = (V_0 \Gamma \omega^4 / 4\pi v_g^3)$,[18] where $V_0$ is the volume per atom, $\omega$ is the phonon frequency, $v_g$ is the phonon group velocity, and $\Gamma$ is the measure of the scattering strength, linearly proportional to the dopant concentration, $N_D$. If the phonon velocity does not change, the increase in the point-defect scattering rate will be defined by the increase in $\Gamma \sim N_D$. However, the reduction of the phonon velocity, $v_g$, will amplify the effect of the doping on the acoustic phonon scattering, and, correspondingly, on thermal conductivity. Similarly, the reduction in the phonon velocity can affect acoustic phonon scattering on dislocation lines, boundaries, and other scattering mechanisms. The thermal boundary resistance, which depends on the phonon density of state mismatch at the interface of two materials can also be altered by the phonon group velocity change.

High thermal conductivity, low TBR, and temperature stability are crucial metrics for UWBG materials.[19–22] To handle high power densities, the device structure should be able to dissipate the heat effectively. Since the variations in the acoustic phonon velocity can affect the bulk and interface thermal conduction, the key questions to address are the following. Is the reduction of the acoustic phonon velocity due to doping a general trend or rather an exception? Does the variation in acoustic phonon group velocity depends on the interplay of the size of the dopant atoms as compared to host atoms, the presence of the strain in the films, or other factors? In this study, we answer these questions by examining the changes in the energy of acoustic and optical phonons as a function of Si doping in a set of AlN films grown on c-plane sapphire substrates. We use Raman spectroscopy to measure the optical phonon frequencies close to the Γ point in the Brillouin zone (BZ). To probe the energy of bulk phonons, we utilize Brillouin-Mandelstam light scattering spectroscopy (BMS).[23] The trends for optical and acoustic phonons with increasing doping concentration are then compared to each other and the dislocation line density variations. No prior studies have been reported for acoustic phonons in Si-doped AlN films, which provides extra motivation for this investigation.

The AlN films for this study were grown using metal-organic chemical vapor deposition (MOCVD) with trimethylaluminum (TMA) and ammonia ($NH_3$) as Al and N precursors, respectively. Initially, 200 nm AlN epilayers were deposited to serve as an AlN/sapphire template, followed by annealing at 1700 °C to reduce the dislocation density. More details about the template preparation can be found elsewhere.[12,24] Subsequently, ~1000 nm of the unintentionally doped (UID) and Si-doped AlN samples were grown at 1100 °C and 20 torr. Silane ($SiH_4$) was used as the Si source, with the Si doping concentration varying between UID and $3\times10^{19}$ cm$^{-3}$. The sample composition and quality were evaluated using Raman spectroscopy (Renishaw InVia) in the standard backscattering configuration with an excitation wavelength of 488 nm focused with a 50× lens and a diffraction grating of 1800 mm$^{-1}$. The laser power was maintained below 1 mW to avoid local heating. No laser-induced changes were observed in Raman peak intensities throughout several measurements. The results of the measurements are presented in Figure 1 (a-c).

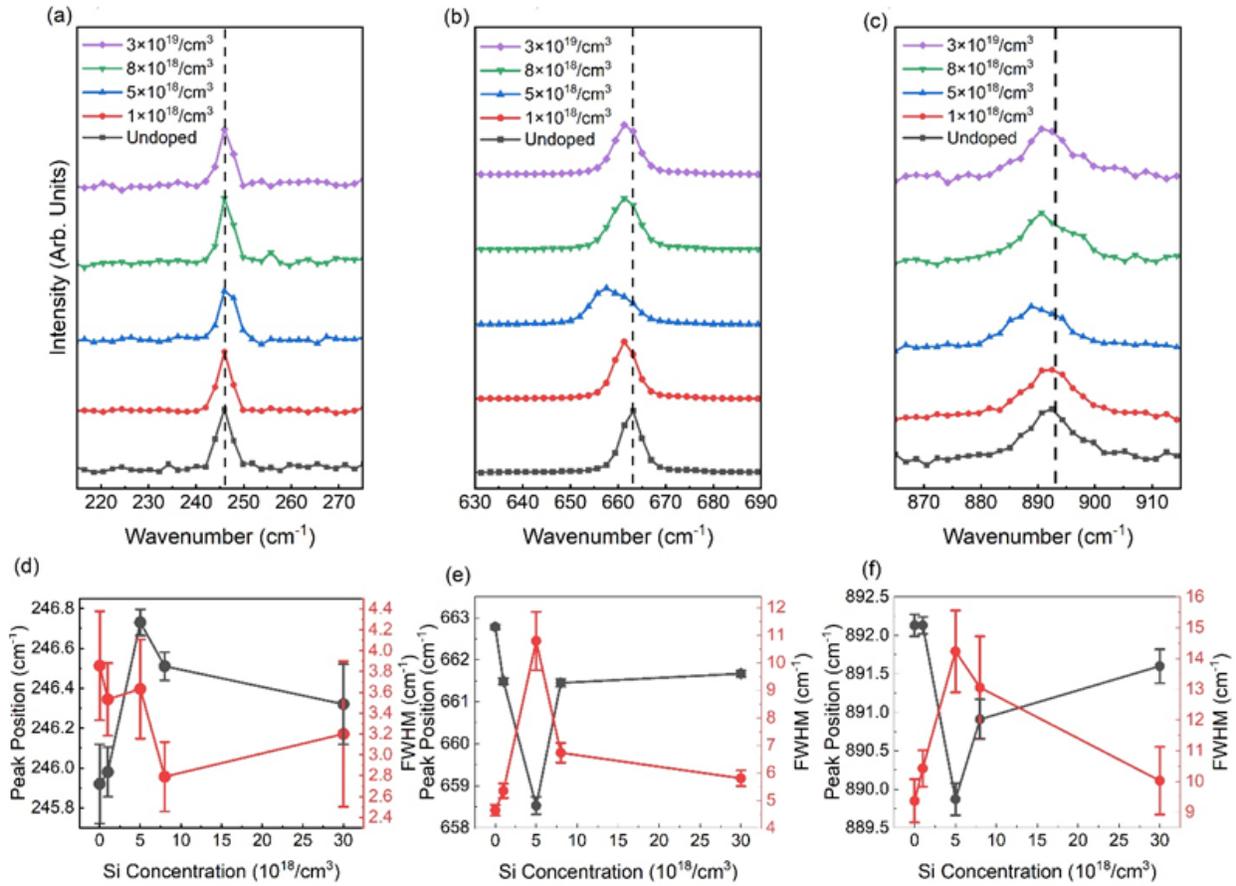

[**Figure 1**: Raman spectra of (a) $E_2^{low}$, (b) $E_2^{high}$, and (c) $A_1$ (LO) phonon modes in the UID and Si-doped AlN films. (d-f) The spectral peak position and full-width-at-half-maximum (FWHM) of the Raman peaks as a function of Si concentration. The vertical dashed lines show the peak position in the UID AlN reference sample.]

AlN belongs to the $C_{6v}^4$ ($P6_3mc$) space group with four Raman active modes: $A_1$, $2E_2$, and $E_1$.[25] All Raman measurements were conducted in the conventional backscattering geometry with the incident light normal to the *c*-plane corresponding to phonon wavevectors along the Γ-A direction in the reciprocal space. The prominent $E_2^{low}$, $E_2^{high}$, and $A_1$ peaks can be seen in Figure 1 (a) - (c), respectively. The experimental Raman data were fitted using Lorentzian functions, from which the peak positions and full-width-at-half-maximum (FWHM) values were extracted for each sample. The dashed lines correspond to the peak positions in the UID AlN sample, which was used as a reference. For the UID AlN, the $E_2^{low}$, $E_2^{high}$, and $A_1$ peaks are located at 245.9 cm⁻¹, 662.8 cm⁻¹, and 892.1 cm⁻¹, respectively. Previous studies on bulk AlN have reported $E_2^{high}$ and $A_1$ peaks at lower wavenumbers, specifically in the range of 657 cm⁻¹ for $E_2^{high}$ and 890 cm⁻¹ for $A_1$. The $E_2^{low}$ mode for bulk AlN is reported to be in the range of 248 cm⁻¹, which is higher than that observed in UID AlN sample.[26,27]

The shift of $E_2^{high}$ and $A_1$ peaks to higher and $E_2^{low}$ to lower wavenumbers in UID samples, compared to the bulk, confirms that the grown film is under residual compressional strain. This observation is consistent with the difference in thermal expansion coefficients between AlN and sapphire, where the lower thermal expansion coefficient of AlN compared to sapphire substrate results in the film experiencing in-plane compressional strain and consequently, tensile strain along the c-lattice direction during cooling.[28] Assuming a biaxial strain in the UID AlN film, the change in the Raman spectral position of each mode, $\delta \omega$ [cm⁻¹], is expressed as, $\delta \omega_i = \chi_i \epsilon_{zz} = [(-C_{33}/C_{13})\alpha + \beta]\epsilon_{zz}$, in which $\alpha$ and $\beta$ are the mode-specific phonon deformation potentials, and $C_{33} = 402$ [GPa] and $C_{13} = 95$ [GPa] are elastic stiffness constants, respectively.[29] The parameter, $\chi_i$ [cm⁻¹], determines the sensitivity of the specific Raman mode to strain along the $c$ direction, $\epsilon_{zz}$. The spectral position of the Raman peaks in bulk and UID AlN film along with the mode-specific deformation potentials and calculated $\chi_i$ are listed in Table I. As seen, $E_2^{high}$ is more sensitive to the strain in the AlN films. Additionally, $E_2^{low}$, has a negative value of $\chi_i$, showing that with tensile strain, $\epsilon_{zz} > 0$, it moves to lower wavenumbers. These are consistent with our experimental observations.

Table I: Raman spectral characteristics of bulk AlN and UID AlN thin films

|  | $E_2^{low}$ [cm⁻¹] | $E_2^{high}$ [cm⁻¹] | $A_1$ [cm⁻¹] | Refs. |
|---|---|---|---|---|
| Bulk | 248 | 657 | 890 | 26 |
| UID | 245.9 | 662.8 | 892.1 | this work |
| $\delta \omega_i$ [cm⁻¹] | -2.1 | +5.8 | +2.1 | - |
| $\alpha$ [cm⁻¹] | 108 | -1145 | -739 | 30 |
| $\beta$ [cm⁻¹] | -237 | -1023 | -737 | 30 |
| $\chi_i$ | -690 | 3786 | 2367 | - |

Generally, doping of the III-nitrides can affect the lattice parameter, and thus the Raman lines, due to a variety of factors including the deformation potential effect, the dopant size effects, residual strain, and the influence of doping on the thermal expansion coefficient.[31] Figures 1 (d-f) show the spectral position and FWHM of the Raman modes for AlN films as a function of Si doping concentration. Both the spectral position and FWHM of Raman modes exhibit a strongly *non-monotonic* trend, indicating a complex dynamic and interplay of several effects. As discussed below, the Raman peak shift can be explained by the interplay of the strain distribution in the grown films, local strain fields generated by the dopant atoms due to their size difference from the host atoms, and its relaxation by forming the dislocation lines.[32]

At and below $5 \times 10^{18}$ cm$^{-3}$, the $E_2^{high}$ and $A_1$ peaks shift to lower wavenumbers compared to that of the UID sample. This trend is the opposite for the $E_2^{low}$ mode owing to its different phonon potential deformation coefficients (see Table I). At $5 \times 10^{18}$ cm$^{-3}$ of Si doping, the spectral position of Raman peaks fell within the range for those reported for bulk AlN, confirming that the initially strained films during the growth are significantly relaxed. Prior studies have reported that addition of Si dopants to AlN lowers the *c*-lattice parameter while slightly decreasing the *a*-lattice parameter.[26] The lattice contraction is likely due to the smaller Si atoms of ~0.42 Å with respect to Al atoms of ~0.51 Å. Incorporating Si atoms into the host matrix causes the previously tensile strain formed in the *c*-lattice direction during the growth to relieve, moving the Raman peaks toward the bulk AlN values. The relaxation of the residual tensile strain can be partially caused by the formation of dislocation lines which will be discussed next. At higher Si concentrations above $5 \times 10^{18}$ cm$^{-3}$, given that the Si—N bonds are shorter than Al—N bonds, local compressional strain starts to form. The contraction in the in-plane direction causes a tensile strain to form in the perpendicular direction, *i.e.,* along the c-lattice direction due to the Poisson effect. This causes the $E_2^{high}$ and $A_1$ modes to blueshift again while causing the $E_2^{low}$ peak move to lower wavenumbers. The non-monotonic evolution of the phonon peaks in AlN with increasing Si doping is consistent with one available prior report.[33]

For all the samples, high-resolution X-ray diffraction (HRXRD) rocking curves were measured along the AlN (0002) peak using an X-ray diffractometer system (Rigaku SmartLab). All samples exhibited good crystallinity with FWHM values less than 0.0356° (~130 arcseconds). We estimated the dislocation density ($D_B$) in the AlN films from FWHM of the rocking curve following the approach and formulas in Ref. 34, $D_B = \beta^2/(4.35b^2)$ where $\beta$ is the FWHM and $b$ is the length of the Burgers vector. The HRXRD and dislocation line density data are presented in Figure 2 (a-b). One can see from Figure 2 that when the doping concentration exceeds $1\times10^{18}$ cm$^{-3}$, the dislocation density is above $10^9$ cm$^{-2}$. For all three samples with doping levels above $5\times10^{18}$ cm$^{-3}$, the defect density is higher than in UID and $1\times10^{18}$ cm$^{-3}$ Si-doped samples. These values are comparable to recent reports on Si-doped AlN.[35–37] The changes in the dislocation line density with the Si doping concentration correlate with the evolution of the optical phonon peaks obtained from Raman spectroscopy.

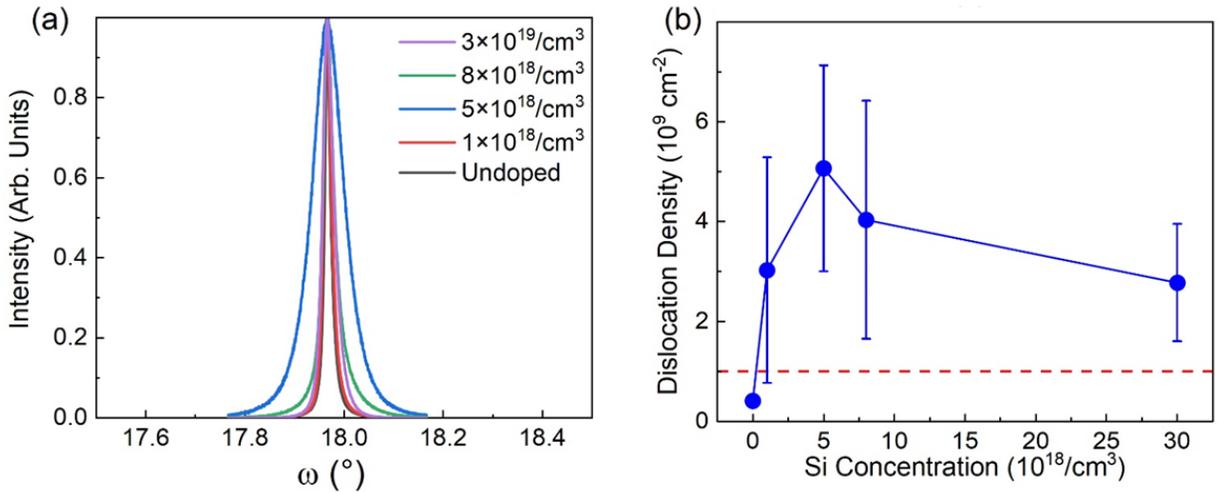

[**Figure 2**: (a) XRD data for (0002) peak of AlN, and (b) the calculated dislocation density based on the FWHM of the rocking curve as a function of Si concentration. For each sample, the data are collected at two different locations and averaged.]

We conducted BMS experiments to investigate the acoustic phonons in the Si-doped AlN films, Interpretation of BMS data requires the knowledge of refractive index, $n$, which was measured using a spectroscopic ellipsometer (ULVAC UNECS-2000). In this instrument, the light source is a broad-spectrum white light with wavelengths in the visible range of ~500-800 nm. The ellipsometer illuminates the sample with unpolarized light. The reflected light is directed into a detector which measures the relative amplitude and phase of the $s$- and $p$-polarized components of the reflected light from the sample. The resulting data is used to model the film thickness and refractive index using Cauchy's equation: $n(\lambda) = A + B/\lambda^2 + C/\lambda^4$.[38] The $A$, $B$, and $C$ coefficients were found by fitting this equation to the measured data. The obtained dependence of the refractive index on the wavelength is presented in Figure 3 (a). The samples show an increase in *n* at low wavelengths, consistent with previous reports of AlN refractive index.[39,40] In Figure 3 (b), we present the refractive index as a function of the doping concentration at $\lambda$=532 nm, which is the wavelength of the excitation laser used in the BMS experiments. The refractive index, *n*, at 532 nm shows a sharp increase at small doping concentrations and a saturation behavior at high doping levels.

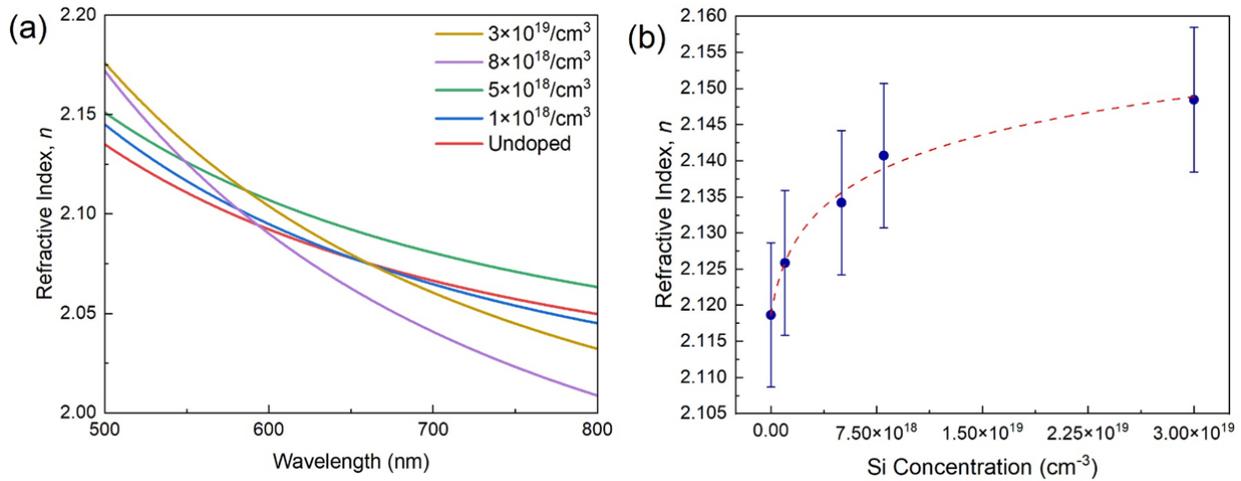

[**Figure 3:** (a) Refractive index, *n*, dispersion of the UID and Si-doped AlN films in the range of 500 nm to 800 nm wavelengths. (b) The index of refraction at 532 nm as a function of Si concentration. The dashed line is for visual clarity.]

The BMS experiments were conducted in the backscattering configuration using a 532 nm laser excitation wavelength at a fixed incidence angle of $\theta = 45°$. The light source was a solid-state diode-pumped continuous-wave laser (Spectra Physics). The laser beam was focused on the sample using an apochromatic objective lens with NA=0.28. The scattered light was collected via the same lens and directed to the high-contrast high-resolution 3 + 3 pass tandem Fabry−Perot interferometer (TFP-2, The Table Stable Ltd., Switzerland) and spectrometer. The mirror spacing of the TFP interferometer was adjusted to 0.8 mm for measuring bulk acoustic phonons. The incident light was *p*-polarized, and the *p*-polarized scattered light was selected for collection. Our BMS procedures are documented in depth in previous reports.[41–43] The spectral positions of the peaks were obtained by fitting the experimental data with individual Lorentzian functions. The BMS data are presented in Figure 4 (a-b). The bulk acoustic phonons modes are probed through the elasto-optic scattering mechanism with the phonon wave vector $q = 4\pi n/\lambda$, where $\lambda = 532$ nm is the excitation wavelength.

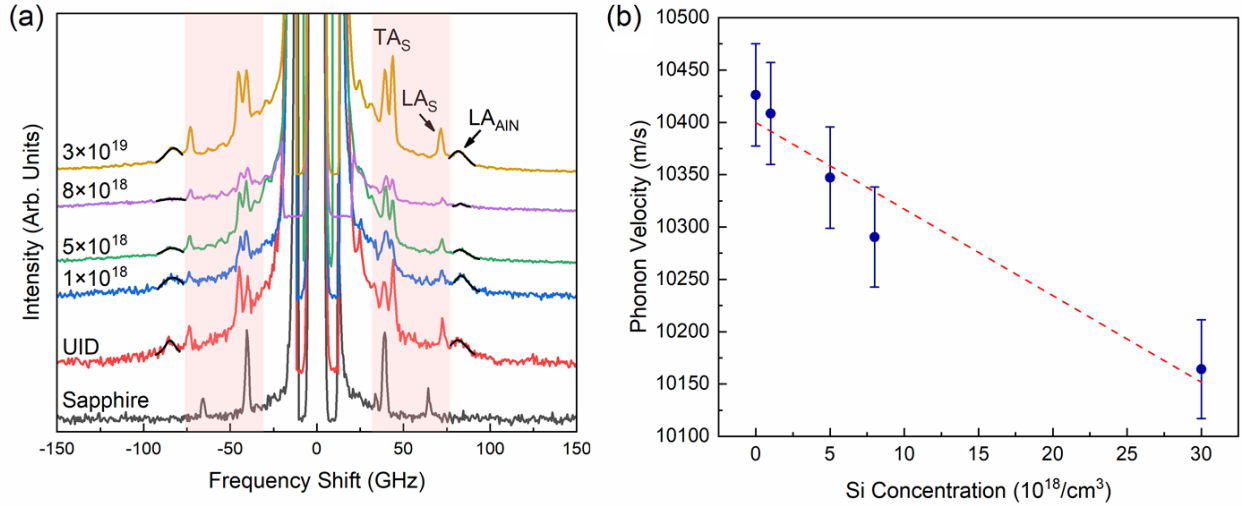

[Figure 4: (a) Brillouin-Mandelstam spectra for the UID and Si-doped AlN samples accumulated at 45° incidence angle. Spectra are translated vertically for visual clarity. The LA$_{AlN}$ peaks were fitted using individual Lorentzian functions (black curves). The subscript 'S' indicates the peaks associated with the sapphire substrate which were highlighted for clarity. (b) Phonon velocity of longitudinal acoustic phonon mode of AlN thin films as a function of Si content.]

In Figure 4 (a), we label the LA phonon in AlN films and sapphire substrate. Since acoustic phonon polarization branches start at zero angular frequency, *i.e.* $\omega(q=0)=0$, one can calculate the phonon group velocity from the formula $v = 2\pi f/q$,[23] where $f$ is the Brillouin shift due to scattering in the material and $q$ is the phonon wave vector. Owing to the linearity of the acoustic phonon dispersion near the BZ center, the group velocity coincides with the phase velocity of the phonons, and it is the same as the sound velocity. Figure 4 (b) shows the dependence of the LA phonon group velocity on the Si doping concentration. The trend for the decreasing phonon velocity with increasing doping is well resolved, and clearly above the experimental uncertainty. The experimental standard error is defined by the ellipsometry instrumentation in the refractive index measurements. The phonon softening with doping is in line with the report for B-doped diamond and Nd-doped sapphire.[15] However, the measured change in the acoustic phonon velocity in our Si-doped AlN thin film sample with the highest doping concentration of $3 \times 10^{19}$ cm$^{-3}$ is relatively small, *i.e.* ~300 m/s, which constitutes about ~3% decrease as compared to the UID reference sample. Below we discuss the possible physical mechanisms behind the measured dependence and explain possible implications of the phonon velocity change for the electronic applications of UWBG semiconductors.

As we found, Si doping in AlN thin films affects optical and acoustic phonons differently. This can be understood, considering that acoustic phonons involve long-wavelength, low-frequency vibrations where atoms move in phase, resulting in relatively modest energy shifts under strain. Optical phonons involve higher-frequency vibrations with atoms moving out of phase within the unit cell, making them more sensitive to changes in bond lengths and angles caused by strain fields. Acoustic phonons are influenced by the overall stiffness and mass density of the crystal lattice, reflecting long-range order. The dopant atoms can introduce lattice distortions and defects that reduce the overall lattice stiffness. In the case of Si-doped AlN, these changes led to a monotonic decrease in the frequency of acoustic phonons as the doping concentration increased (see Figure 4 (b)). The sensitivity of acoustic and optical phonons to various dopants and defects depends on the specific characteristics of a given material system. For example, AlN has polar covalence bonds while diamond, another important UWBG semiconductor, is non-polar. In polar materials, doping may produce a stronger impact on optical phonons, and correspondingly on Raman peaks; in non-polar materials, the doping may affect stronger acoustic phonons and Brillouin peaks. The electronegativity difference between Al and N is ~1.43 while for Si and N it is 1.14, resulting in the corresponding changes in relevant Raman peaks. The observed different sensitivity of the optical and acoustic phonons to dopants and defects can be used to gain more information about the material using two complementary techniques, Raman and BMS spectroscopies.

Let us now consider the relative strength of the effect of dopants on acoustic phonons. As one can see from Figure 4 (b), the acoustic phonon velocity decreases from 10425 m/s in the UID AlN reference sample to 10160 m/s in the thin film with a Si dopant concentration of $3\times10^{19}$ $cm^{-3}$. This is a velocity reduction of ~2.5 %. In the B-doped diamond, the LA phonon experienced a ~2.4% velocity reduction at $10^{17}$ $cm^{-3}$ doping concentration while AlN films had only a ~0.3% velocity reduction for bulk acoustic phonons at $10^{18}$ $cm^{-3}$ Si doping. The velocity of the surface acoustic phonon was reduced up to ~15% in the diamond films with the B concentration of $3\times10^{20}$ cm$^{-3}$ (see Ref [15]). This indicates that Si dopants in AlN films produce a weaker effect on the acoustic phonon velocity. One can consider the relatively small change in the velocity as beneficial for the thermal management of devices based on Si-doped AlN. As mentioned above the acoustic phonon scattering on point defects, *i.e.* dopant atoms, will only be defined by the concentration and not amplified by the reduced phonon velocity. The same is true about acoustic phonon scattering on dislocations. The phonon scattering rate on the dislocation line cores is given by $\tau_C^{-1} = (N_L V_0^{4/3} \omega^3 / v_g^2)$,[4] where $N_L$ is the concentration of dislocation lines. One can see that the reduction in the phonon group velocity would result in an enhanced scattering rate.

The effect of the acoustic phonon velocity change on the thermal boundary resistance, $R_B$, can be even stronger. The value of $R_B$ depends on the phonon transmission probability, which is defined as $\alpha_{1\to 2} = \sum_j v_{2,j}^{-2} / (\sum_j v_{1,j}^{-2} + \sum_j v_{2,j}^{-2})$.[44–46] As seen, $\alpha$ depends on the phonon group velocity, $v_{i,j}$, of adjoining materials; any changes in the phonon velocity will affect $\alpha$, and thus, $R_B$. The measured acoustic phonon velocity can be used for accurate calculation of the thermal boundary resistance, $R_B$. If the effect of the dopant atoms in certain material systems on the acoustic phonon velocity is strong, it does not necessarily mean a detrimental effect on thermal management. One can envision the possibility of using the dopant introduction into the material with a higher phonon velocity to reduce the acoustic phonon velocity ratio for the layers in heterostructures with the corresponding reduction in $R_B$.

In conclusion, we investigated the acoustic and optical phonons in Si-doped AlN thin films on sapphire substrates. The optical phonon frequencies reveal non-monotonic changes, reflective of the variations in the thin-film strain and dislocation densities with the addition of Si dopant atoms. The acoustic phonon velocity decreases monotonically with increasing Si dopant concentration, but the overall change is relatively small. The knowledge of the acoustic phonon velocities can be used for the optimization of the UWBG semiconductor heterostructures and for minimizing the thermal boundary resistance of high-power devices.

**Acknowledgment**

The work at UCLA and ASU was supported by ULTRA, an Energy Frontier Research Center (EFRC) funded by the U.S. Department of Energy, Office of Science, Basic Energy Sciences under Award # DE-SC0021230. A.A.B. and F.K. acknowledge the support of the National Science Foundation (NSF) via a Major Research Instrument (MRI) DMR Project No. 2019056 entitled "Development of a Cryogenic Integrated Micro-Raman-Brillouin-Mandelstam Spectrometer." H.F., D.H.M. and B.D. acknowledge the support of the National Science Foundation (NSF) under Award No. ECCS-2338604.

**Conflict of Interest**

The authors declare no conflict of interest.

**Author Contributions**

A.A.B. and H.F. conceived the idea. A.A.B. coordinated the project, and led the data analysis and manuscript preparation; D.W. conducted Brillouin spectroscopy and contributed to data analysis; E.G. and X.F. conducted Raman spectroscopy and contributed to data analysis; J.T. conducted refractive index measurements and contributed to data analysis; F.K. contributed to Raman and Brillouin data analysis; D.H.M. synthesized the samples, conducted microscopy and XRD studies; H.F. supervised sample preparation and contributed to data analysis. D.W. wrote the initial version of the manuscript. All authors contributed to the manuscript preparation.

**The Data Availability Statement**

The data in support of the findings of this study are available from the corresponding author upon reasonable request.

# REFERENCES

[1] J.Y. Tsao, S. Chowdhury, M.A. Hollis, D. Jena, N.M. Johnson, K.A. Jones, R.J. Kaplar, S. Rajan, C.G. Van de Walle, E. Bellotti, C.L. Chua, R. Collazo, M.E. Coltrin, J.A. Cooper, K.R. Evans, S. Graham, T.A. Grotjohn, E.R. Heller, M. Higashiwaki, M.S. Islam, P.W. Juodawlkis, M.A. Khan, A.D. Koehler, J.H. Leach, U.K. Mishra, R.J. Nemanich, R.C.N. Pilawa-Podgurski, J.B. Shealy, Z. Sitar, M.J. Tadjer, A.F. Witulski, M. Wraback, and J.A. Simmons, "Ultrawide-Bandgap Semiconductors: Research Opportunities and Challenges," Adv Electron Mater **4**(1), (2018).

[2] M.H. Wong, O. Bierwagen, R.J. Kaplar, and H. Umezawa, "Ultrawide-bandgap semiconductors: An overview," J Mater Res **36**(23), 4601–4615 (2021).

[3] B.A. Klein, A.A. Allerman, and A.M. Armstrong, "Al-rich AlGaN high electron mobility transistor gate metallization study up to 600 °C in air," Appl Phys Lett **124**(10), (2024).

[4] C. Peterson, F. Alema, A. Bhattacharyya, Z. Ling, S. Roy, A. Osinsky, and S. Krishnamoorthy, "Kilovolt-class β-Ga2O3 MOSFETs on 1-in. bulk substrates," Appl Phys Lett **124**(8), (2024).

[5] Z. Liang, H. Du, Y. Yuan, Q. Wang, J. Kang, H. Zhou, J. Zhang, Y. Hao, X. Wang, and G. Zhang, "Ultra-thin AlGaN/GaN HFET with a high breakdown voltage on sapphire substrates," Appl Phys Lett **119**(25), (2021).

[6] M. Wohlfahrt, M.J. Uren, Y. Yin, K.B. Lee, and M. Kuball, "Vertical field inhomogeneity associated with threading dislocations in GaN high electron mobility transistor epitaxial stacks," Appl Phys Lett **119**(24), (2021).

[7] R. Yu, G. Liu, G. Wang, C. Chen, M. Xu, H. Zhou, T. Wang, J. Yu, G. Zhao, and L. Zhang, "Ultrawide-bandgap semiconductor AlN crystals: growth and applications," J Mater Chem C Mater **9**(6), 1852–1873 (2021).

[8] C. Bacaksiz, H. Sahin, H.D. Ozaydin, S. Horzum, R.T. Senger, and F.M. Peeters, "Hexagonal AlN: Dimensional-crossover-driven band-gap transition," Phys Rev B Condens Matter Mater Phys **91**(8), (2015).

[9] W. Liu, and A.A. Balandin, "Thermal conduction in AlxGa1-xN alloys and thin films," J Appl Phys **97**(7), (2005).

[10] M.S. Bin Hoque, Y.R. Koh, J.L. Braun, A. Mamun, Z. Liu, K. Huynh, M.E. Liao, K. Hussain, Z. Cheng, E.R. Hoglund, D.H. Olson, J.A. Tomko, K. Aryana, R. Galib, J.T. Gaskins, M.M.M. Elahi, Z.C. Leseman, J.M. Howe, T. Luo, S. Graham, M.S. Goorsky, A. Khan, and P.E. Hopkins, "High In-Plane Thermal Conductivity of Aluminum Nitride Thin Films," ACS Nano **15**(6), 9588–9599 (2021).

[11] M. Khizar, Z.Y. Fan, K.H. Kim, J.Y. Lin, and H.X. Jiang, "Nitride deep-ultraviolet light-emitting diodes with microlens array," Appl Phys Lett **86**(17), 1–3 (2005).

[12] P. Bagheri, C. Quiñones-Garcia, D. Khachariya, S. Rathkanthiwar, P. Reddy, R. Kirste, S. Mita, J. Tweedie, R. Collazo, and Z. Sitar, "High electron mobility in AlN:Si by point and extended defect management," J Appl Phys **132**(18), (2022).

[13] Y. Taniyasu, M. Kasu, and T. Makimoto, "Field emission properties of heavily Si-doped AlN in triode-type display structure," Appl Phys Lett **84**(12), 2115–2117 (2004).

[14] Y. Taniyasu, M. Kasu, and T. Makimoto, "Electrical conduction properties of n-type Si-doped AlN with high electron mobility (>100 cm2 V-1 s-1)," Appl Phys Lett **85**(20), 4672–4674 (2004).

[15] E. Guzman, F. Kargar, F. Angeles, R.V. Meidanshahi, T. Grotjohn, A. Hardy, M. Muehle, R.B. Wilson, S.M. Goodnick, and A.A. Balandin, "Effects of Boron Doping on the Bulk and Surface Acoustic Phonons in Single-Crystal Diamond," ACS Appl Mater Interfaces **14**(37), 42223–42231 (2022).

[16] A. Balandin, and K.L. Wang, "Significant decrease of the lattice thermal conductivity due to phonon confinement in a free-standing semiconductor quantum well," Phys. Rev. B **58**(3), 1544–1549 (1998).

[17] F. Kargar, S. Ramirez, B. Debnath, H. Malekpour, R.K. Lake, and A.A. Balandin, "Acoustic phonon spectrum and thermal transport in nanoporous alumina arrays," Appl Phys Lett **107**(17), (2015).

[18] P.G. Klemens, *Heat Conduction in Solids by Phonons* (1993).

[19] T. Feng, H. Zhou, Z. Cheng, L.S. Larkin, and M.R. Neupane, "A Critical Review of Thermal Boundary Conductance across Wide and Ultrawide Bandgap Semiconductor Interfaces," ACS Appl Mater Interfaces **15**(25), 29655–29673 (2023).

[20] S. Khan, F. Angeles, J. Wright, S. Vishwakarma, V.H. Ortiz, E. Guzman, F. Kargar, A.A. Balandin, D.J. Smith, D. Jena, H.G. Xing, and R. Wilson, "Properties for Thermally Conductive Interfaces with Wide Band Gap Materials," ACS Appl Mater Interfaces **14**(31), 36178–36188 (2022).

[21] H.N. Masten, J.S. Lundh, T.I. Feygelson, K. Sasaki, Z. Cheng, J.A. Spencer, P.Y. Liao, J.K. Hite, D.J. Pennachio, A.G. Jacobs, M.A. Mastro, B.N. Feigelson, A. Kuramata, P. Ye, S. Graham, B.B. Pate, K.D. Hobart, T.J. Anderson, and M.J. Tadjer, "Reduced temperature in lateral (AlxGa1−x)2O3/Ga2O3 heterojunction field effect transistor capped with nanocrystalline diamond," Appl Phys Lett **124**(15), (2024).

[22] J. Niroula, Q. Xie, N.S. Rajput, P.K. Darmawi-Iskandar, S.I. Rahman, S. Luo, R.H. Palash, B. Sikder, M. Yuan, P. Yadav, G.K. Micale, N. Chowdhury, Y. Zhao, S. Rajan, and T. Palacios, "High temperature stability of regrown and alloyed Ohmic contacts to AlGaN/GaN heterostructure up to 500 °C," Appl Phys Lett **124**(20), (2024).

[23] F. Kargar, and A.A. Balandin, "Advances in Brillouin-Mandelstam light-scattering spectroscopy," Nat. Photon **15**, 720–731 (2021).

[24] H. Miyake, G. Nishio, S. Suzuki, K. Hiramatsu, H. Fukuyama, J. Kaur, and N. Kuwano, "Annealing of an AlN buffer layer in N2-CO for growth of a high-quality AlN film on sapphire," Applied Physics Express **9**(2), (2016).

[25] K. Ishioka, K. Kato, N. Ohashi, H. Haneda, M. Kitajima, and H. Petek, "The effect of n- and p-type doping on coherent phonons in GaN," Journal of Physics Condensed Matter **25**(20), (2013).

[26] V.Y. Davydov, Y.E. Kitaev, I.N. Goncharuk, A.N. Smirnov, J. Graul, O. Semchinova, D. Uffmann, M.B. Smirnov, A.P. Mirgorodsky, and R.A. Evarestov, *Phonon Dispersion and Raman Scattering in Hexagonal GaN and AlN* (n.d.).

[27] L.E. McNeil, M. Grimsditch, and R.H. French, "Vibrational Spectroscopy of Aluminum Nitride," Journal of the American Ceramic Society **76**(5), 1132–1136 (1993).

[28] T. Kallel, M. Dammak, J. Wang, and W.M. Jadwisienczak, "Raman characterization and stress analysis of AlN:Er3+ epilayers grown on sapphire and silicon substrates," Materials Science and Engineering: B **187**, 46–52 (2014).

[29] K. Shojiki, Y. Hayashi, K. Uesugi, and H. Miyake, "Local and anisotropic strain in AlN film on sapphire observed by Raman scattering spectroscopy," Jpn J Appl Phys **58**(SC), (2019).

[30] H. Harima, *To Cite This Article: Hiroshi Harima* (2002).

[31] M.A. Moram, and M.E. Vickers, "X-ray diffraction of III-nitrides," Reports on Progress in Physics **72**(3), (2009).

[32] S.B. Thapa, J. Hertkorn, F. Scholz, G.M. Prinz, M. Feneberg, M. Schirra, K. Thonke, R. Sauer, J. Biskupek, and U. Kaiser, "MOVPE growth of high quality AlN layers and effects of Si doping," in *Physica Status Solidi (C) Current Topics in Solid State Physics*, (2008), pp. 1774–1776.

[33] G.M. Prinz, M. Feneberg, M. Schirra, R. Sauer, K. Thonke, S.B. Thapa, and F. Scholz, "Silicon-doping induced strain of aln layers: A comparative luminescence and Raman study," Physica Status Solidi - Rapid Research Letters **2**(5), 215–217 (2008).

[34] M.A. Moram, and M.E. Vickers, "X-ray diffraction of III-nitrides," Reports on Progress in Physics **72**(3), (2009).

[35] A. Nakajima, Y. Furukawa, H. Yokoya, S. Yamaguchi, and H. Yonezu, "Growth of low-dislocation-density Al N under Ga irradiation," Japanese Journal of Applied Physics, Part 1: Regular Papers and Short Notes and Review Papers **45**(4 A), 2422–2425 (2006).

[36] M. Nemoz, R. Dagher, S. Matta, A. Michon, P. Vennéguès, and J. Brault, "Dislocation densities reduction in MBE-grown AlN thin films by high-temperature annealing," J Cryst Growth **461**, 10–15 (2017).

[37] K.M. Pürlü, M.N. Koçak, G. Yolcu, İ. Perkitel, İ. Altuntaş, and I. Demir, "Growth and characterization of PALE Si-doped AlN on sapphire substrate by MOVPE," Mater Sci Semicond Process **142**, (2022).

[38] A.-L. Cauchy, "Sur la réfraction et la réflexion de la lumière," Bulletin Des Sciences Mathématiques **14**(6), (1830).

[39] J.F. Muth, J.D. Brown, M.A.L. Johnson, Z. Yu, R.M. Kolbas, J.W. Cook, and J.F. Schetzina, "Absorption coefficient and refractive index of GaN, AlN, and AlGaN Alloys," MRS Internet J. Nitride Semicond. Res **4S1**(G5.2), (1999).

[40] N.M. Figueiredo, F. Vaz, L. Cunha, S.E. Rodil, and A. Cavaleiro, "Structural, chemical, optical and mechanical properties of Au doped AlN sputtered coatings," Surf Coat Technol **255**, 130–139 (2014).

[41] E. Guzman, F. Kargar, A. Patel, S. Vishwakarma, D. Wright, R.B. Wilson, D.J. Smith, R.J. Nemanich, and A.A. Balandin, "Optical and acoustic phonons in turbostratic and cubic boron nitride thin films on diamond substrates," Diam Relat Mater **140**, (2023).

[42] F. Kargar, B. Debnath, J.P. Kakko, A. Saÿnätjoki, H. Lipsanen, D.L. Nika, R.K. Lake, and A.A. Balandin, "Direct observation of confined acoustic phonon polarization branches in free-standing semiconductor nanowires," Nat Commun **7**, (2016).

[43] D. Wright, Z. Ebrahim Nataj, E. Guzman, J. Polster, M. Bouman, F. Kargar, and A.A. Balandin, "Acoustic and optical phonons in quasi-two-dimensional MPS3 antiferromagnetic semiconductors," Appl Phys Lett **124**(16), (2024).

[44] E.T. Swartz, and R.O. Pohl, "Thermal boundary resistance," Rev. Mod. Phys. **61**(3), 605–668 (1989).

[45] S.C. Lee, and G.R. Cunnington, "Conduction and radiation heat transfer in high-porosity fiber thermal insulation," J Thermophys Heat Trans **14**(2), 121–136 (2000).

[46] L. De Bellis, P.E. Phelan, and R.S. Prasher, "Variations of acoustic and diffuse mismatch models in predicting thermal-boundary resistance," J Thermophys Heat Trans **14**(2), 144–150 (2000).